\documentclass[english,12pt,a4paper]{article}
\usepackage[T1]{fontenc}
\usepackage{graphicx}
\usepackage{mathtools}
\usepackage{amssymb}
\usepackage{amsthm}
\usepackage{babel}
\usepackage{hyperref}
\usepackage{geometry} %flexible way to change page orientation and margins
\title{Nonlinear Hall effect in the stationary cylinder with a radial heat flux}
\date{}
\newcommand{\footremember}[2]{%
	\footnote{#2}
	\newcounter{#1}
	\setcounter{#1}{\value{footnote}}%
}
\newcommand{\footrecall}[1]{%
	\footnotemark[\value{#1}]%
} 
\author{%
	G. S. Bisnovatyi-Kogan\footremember{one}{Space Research Institute Russian Academy of Sciences}, \footremember{two}{NRNUM MEPhI}%
	\and M. V. Glushikhina\footrecall{one}
}
\begin{document}
	
	\maketitle
	
	\begin{abstract}
		A conducting cylinder with a uniform magnetic field along its axis
		and radial temperature gradient is considered at the stationary state.
		At large temperature gradients
		the azimuthal Hall electrical current
		creates the axial magnetic field which strength may be comparable with the original one.
		It is shown, that the  magnetic field,
		generated by the  azimuthal Hall current, leads to the decrease of magnetic field  originated by external sources,  and
		this suppression increases with increase of the  electromotive force, connected with a thermodiffusion.  Obtained results can help to investigate
		influence of the Hall current on the coupled magneto-thermal evolution  of
		magnetic and electric fields in neutron stars, white dwarfs, and, possibly, in a laboratory facilities.
	\end{abstract}

	\section{Introduction}
	
	X-ray observations of some isolated neutron stars (NS) show  periodic variabilities of their thermal emission, indicating to the anisotropic temperature distribution.
	One can say that the geometry of the magnetic field in the interior of a NS leaves an observable imprint on the surface, potentially allows us to study the internal structure of the magnetic field through modeling of the spectra and pulse profile of thermally emitting NSs.
	Transport coefficients determining a heat flux and diffusion (electrical current)
	in plasma have a tensor structure in presence of a magnetic field. It
	means, that a direction of the heat and diffusion fluxes do not coincide with a direction of corresponding vectors of electrical field $\textit E$,
	and temperature gradient ${\nabla} \textit{T}$, responsible for these fluxes formation.  Difference of transport coefficients is related to differences of  fluxes along and perpendicular to the magnetic field direction.
	A drift motion of  charged particles
	~\cite{Alfven-Faelthammar-1963},
	in the direction perpendicular to the
	plane to which both $\textit E$ and $\textit B$ belong, determines the electrical current flux ${\textit j}_H$ along this perpendicular, which is called
	as {\it Hall current}. Same property is characteristic for the electronic heat flux current ${\textit Q}_H$.
	Influence of Hall current on magnetized plasmas behaviour in laboratory con\-di\-ti\-ons  was studied by
	\cite{Frucht-1992}, \cite{Gomberoff-Fruchtman-1993},
	\cite{Gomez-Mahajan-Dmitruk-2008}.
	
	In astrophysical  objects an effect of Hall currents on the magnetic field geometry  was studied in the work  \cite{Gold-1992} where they  analysed magnetic field decay in an isolated neutron star.  In work \cite{Gourgouliatos-Cumming-2015} braking index measurements of young radio pulsars explained by influence of magnetic field evolution in the neutron star crust due to Hall drift. In  the work of \cite{Gourgouliatos-Wood-2016} three-dimensional simulations were  presented for magnetic field in magnetar crusts.
	
	In the paper \cite{Vigano-Garcia-Pons-2021} performed a simulation of temperature and magnetic field evolution of neutron stars with coupled ohmic, hall and ambipolar effects; Pons et al.   \cite{Pons-Vigano-2016} reviewed  theo\-re\-ti\-cal and numerical research of neutron stars magneto-thermal evolution, supp\-le\-men\-ted with detailed calculations of microphysical properties.
	
	Determination of transport coefficient tensors from solution of
	Boltzmann ki\-ne\-tic equation was described in the classical book of  \cite{Chapman-1952}.
	
	Application to laboratory and astrophysical plasma of this theory, and calculations of transport coefficients by method described in  the book of \cite{Chapman-1952}, are performed by \cite{Brag-1957}. In the papers of \cite{BK-Glush-2018},  \cite{BK-Glush-2018a},
	\cite{Glu-2020}
	such calculations have been performed for wider region of parameters, including the case of strongly degenerate electrons.
	
	The heat and diffusion fluxes in plasma are governed by diffusion vector $\textit{\bf d}$ and temperature gradient vector ${\bf\nabla} \textit T$.
	In presence of a magnetic field $\textit{\bf B}$ the connection of fluxes with these vectors has a tensor structure.
	A part of the electrical current vector $\textit 
	{\bf j}$ is connected with the electrical field  
	vector $\textit{\bf E}$, consisting the main part 
	of the diffusion vector $\textit{\bf d}$, by 
	electrical conductivity tensor 
	{$\overleftrightarrow\sigma_E $}. Another part of 
	$\textit{\bf j}$ is connected with the temperature 
	gradient vector ${\bf\nabla} \textit T$ by a  
	tensor {$\overleftrightarrow\sigma_T$}.
	
	In a non-degenerate non-magnetized plasma, the scalar electron thermo-diffusion coefficient $\sigma_T$ is connected with the scalar heat conductivity coefficient $\tilde\lambda_T$, related to ${\bf\nabla} \textit T$, as \cite{BK-Glush-2018,Glu-2020}
	
	\begin{equation}
		\sigma_T \approx \frac{3\,e\, \tilde\lambda_T}{20\,kT}.
		\label{eq1}
	\end{equation}
	This relation becomes exact in the Lorenz gas approximation \cite{BK-2001}.
	
	In  following, we discuss behaviour of magnetic field in stationary state, generated by the azimuthal Hall current, produced by temperature gradient only. Obtained results can be used for  evaluating temperature distribution on the neutron star's surface, modeling structure of magnetic field on the surface and in the crust as well as for studying magnetic and electric field distribution in plasma in laboratory conditions.
	
	\section{Magnetic fields, electromotive force, and elect\-ri\-cal currents in a conducting cylinder	\label{2}}

	In the paper of \cite{BK-Glush-2018} the following general relations in Cartesian coordinates were written for the  four kinetic coefficients, namely heat conductivity ($\lambda_{ij}$), diffusion ($\eta_{ij}$), thermodiffusion ($\mu_{ij}$) and diffusional thermal effect ($\nu_{ij}$)  of electrons in non-degenerate non-relativistic plasma,  that depends on magnetic field $B_i$, concentration of electrons $n_e$, electric field $E_i$, temperature $T$ and mass-average velocity $c_{0k}$:
	\begin{eqnarray}
		\label{q_i}
		q_i=q_{i}^{(T)}+q_{i}^{(D)}=
		-\left(\lambda^{(1)}\delta_{ij}-\lambda^{(2)}\varepsilon_{ijk}B_k+\lambda^{(3)}B_i B_j\right)\frac{\partial T}{\partial x_j} \nonumber\\
		-n_e \left(\nu^{(1)}\delta_{ij}-\nu^{(2)}\varepsilon_{ijk}B_k+\nu^{(3)}B_i B_j\right)d_j,
	\end{eqnarray}.
	
	\begin{eqnarray}
		\label{v_i}
		\langle v_{i} \rangle=\langle v_{i}^{(D)} \rangle + \langle
		v_{i}^{(T)} \rangle \nonumber\\
		=-n_e \left(\eta^{(1)}\delta_{ij}
		-\eta^{(2)}\varepsilon_{ijk}B_k+\eta^{(3)}B_i B_j\right)d_j \\
		-\left(\mu^{(1)}\delta_{ij}-\mu^{(2)}\varepsilon_{ijk}B_k
		+\mu^{(3)}B_i B_j\right)\frac{\partial T}{\partial x_j},\nonumber
	\end{eqnarray}
	
	\begin{eqnarray}
		\label{d_i}
		d_{i} = \frac{\rho_{N}}{\rho} \frac{ \partial \ln P_{e}}{\partial x_{i}}
		- \frac{\rho_{e}}{P_{e}}\frac{1}{\rho}\frac{\partial P_{N}}{\partial x_{i}}
		+\frac{e}{kT}(E_i+\frac{1}{c} \varepsilon _{ikl} c_{0k} B_{l}).
	\end{eqnarray}
	The indices (T) and (D) correspond to the heat flux
	$q_i$, and diffusion velocity $\langle v_{i} \rangle$ of electrons, determined by temperature gradient $\partial T/\partial x_j$, and diffusion vector $d_j$, respectively.
	
\begin{figure}
		\begin{center}
			\includegraphics[width=0.42\textwidth]{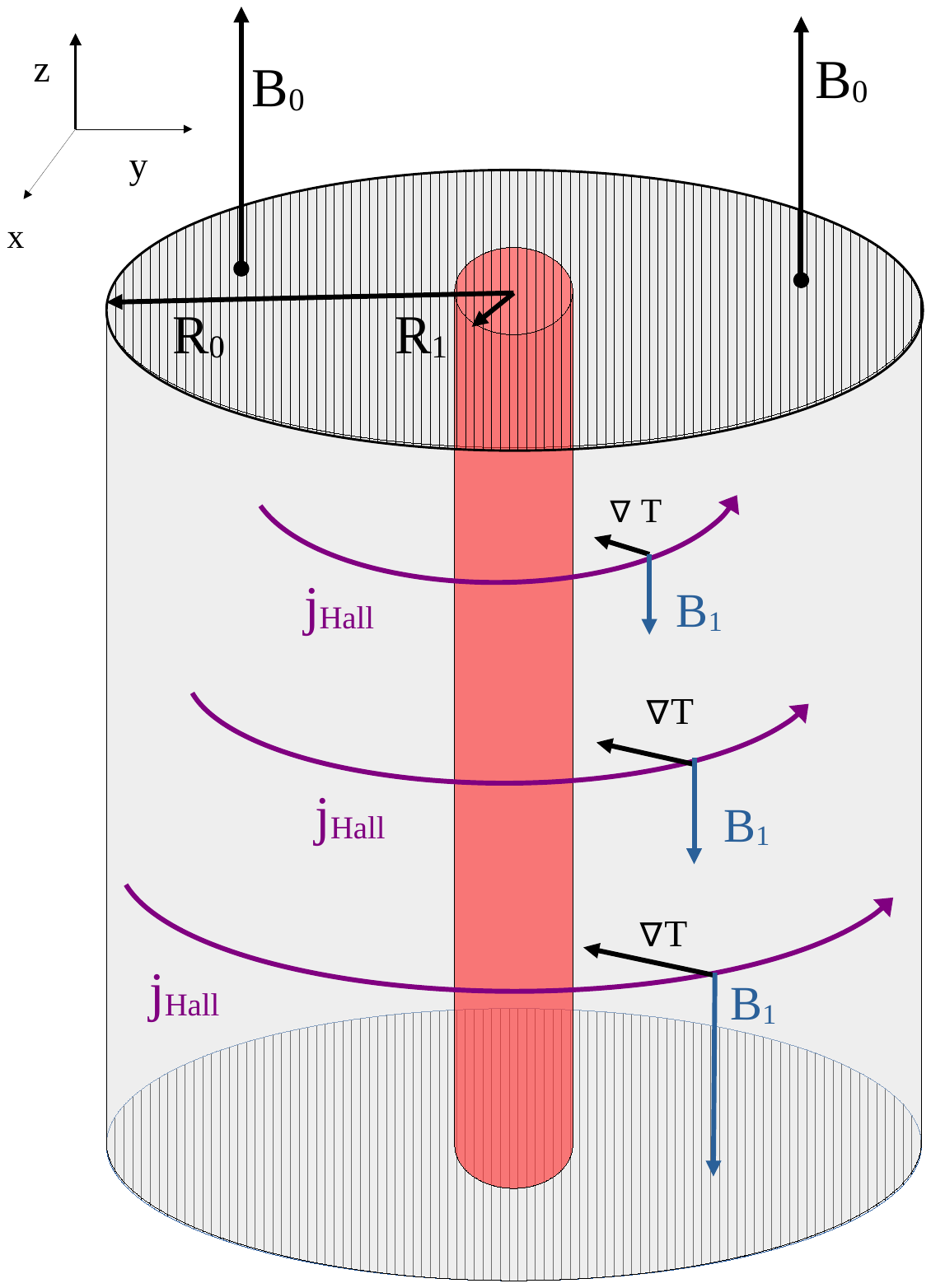}
		\end{center}	
		\caption{Conducting cylinder with Hall current $j_{Hall}$, depending on the magnitude of the radial temperature gradient, and external constant magnetic field $B_0$ along its axis. The induced magnetic field $B_1$ is determined by the Hall current. $R_1$ is the radius of the  central heated region with constant temperature $T_0$. Toroidal region, coloured in gray, contains Hall current and associated magnetic field, which has an opposite direction to the external field $B_0$, decreasing the resulting field along the cylinder. }\label{cylinder}
	\end{figure}

\begin{figure}
		\begin{center}
			\includegraphics[width=0.42\textwidth]{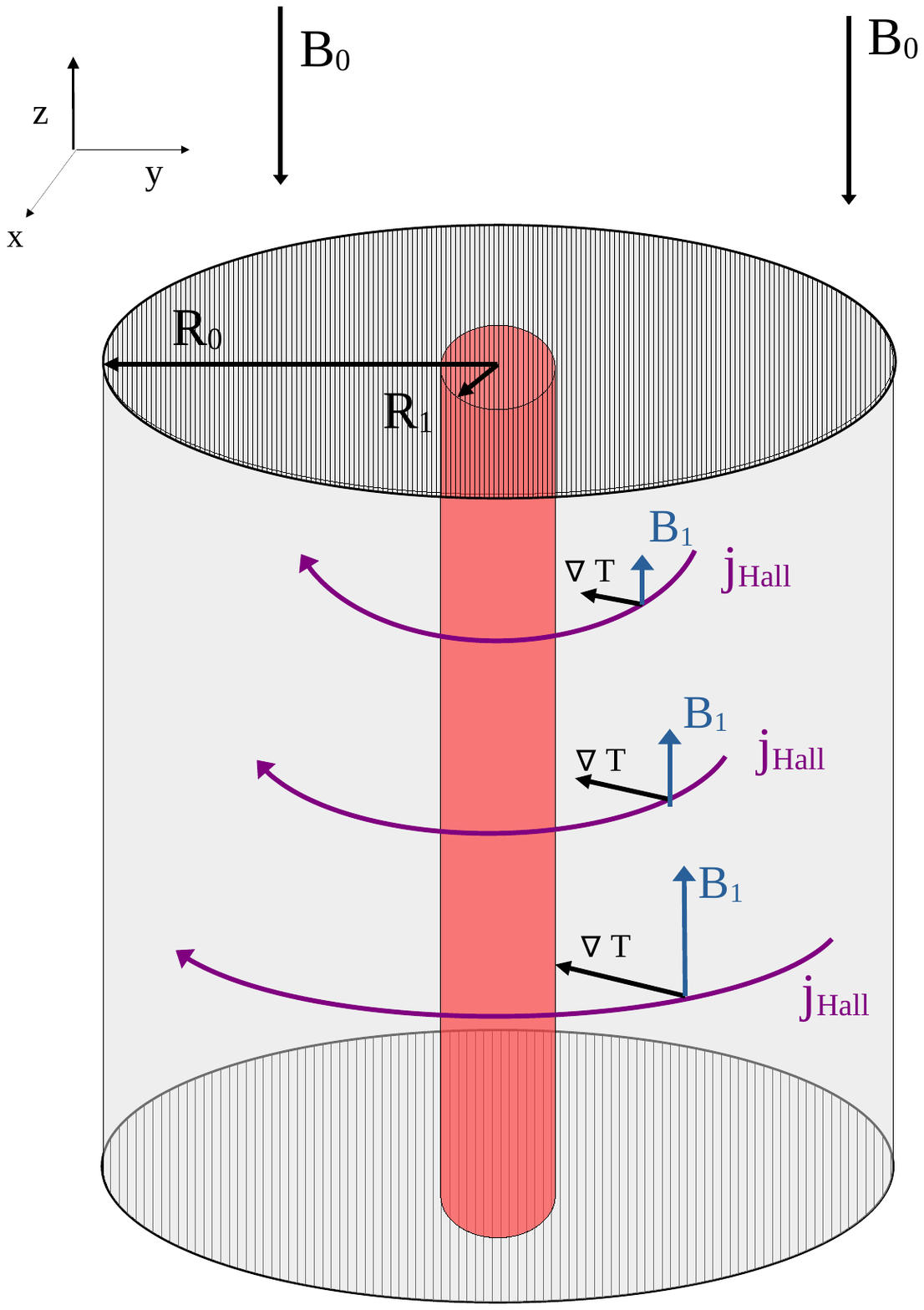}
		\end{center}	
		\caption{The same  cylinder as in Fig.\ref{cylinder},  with opposite direction of the constant magnetic field $B_0$. We see, that the magnetic field $B_1$, induced by Hall currents $j_{Hall}$ is again opposite to the direction of $B_0$. Therefore the resulting magnetic field decreases, for any direction of the magnetic field $B_0$.}\label{cylinder1}
	\end{figure}

	Here  $P_e$ is  the electron pressure, $P_N$ is the ion pressure, $\rho$ is the density, defined as $\rho = m_N n_N$,  $n_N$ is concentration of ions. The tensor kinetic coefficients
	$\lambda^{(i)}$, $\mu^{(i)}$, $\eta^{(i)}$ and $\nu^{(i)}$ determine the heat and diffusion fluxes
	in the following directions. The upper indices $^{(1)}$
	determine the above mentioned fluxes along the temperature gradient
	$\partial T/\partial x_i$, or diffusion vector $d_i$.
	The upper indices $^{(3)}$ are related to the direction along the magnetic field; and the upper indices $^{(2)}$ determine fluxes perpendicular to the plane defined by the magnetic field vector $B_i$ and any of the vectors $\partial T/\partial x_i$ or $d_i$. These last fluxes are referred to as the Hall ones, $q_{Hall}$ and $j_{Hall}$.
	We consider here terms in the heat flux  and the electrical current produced by the temperature gradient only, so equations (\ref{q_i}), (\ref{v_i}) can be written as

	\begin{eqnarray}
		\label{q_ii}
		q_i=q_{i}^{(T)}=	-\left(\lambda^{(1)}\delta_{ij}-\lambda^{(2)}\varepsilon_{ijk}B_k+\lambda^{(3)}B_i B_j\right)\frac{\partial T}{\partial x_j},
	\end{eqnarray}
	
	\begin{eqnarray}
		\label{v_ii}
		\langle v_{i} \rangle=\langle v_{i}^{(T)} \rangle =
		-\left(\mu^{(1)}\delta_{ij}-\mu^{(2)}\varepsilon_{ijk}B_k+\mu^{(3)}B_i B_j\right)\frac{\partial T}{\partial x_j}.
	\end{eqnarray}
	Let us consider a plasma cylinder (see Figs.\ref{cylinder}, \ref{cylinder1}) with a uniform magnetic field $B$ along $z$ axis, a temperature gradient vector along the radius. 
	In the case of a cylinder symmetry $\frac{\partial}{\partial z}=\frac{\partial}{\partial\phi}=0$,  the only non-zero parameters are
	$q_r,\,\,q_\phi,\,\,j_r,\,\,j_\phi,\,\,B_z$.
	Using the definition of the electrical current
	\begin{equation}
		\label{ji}
		j_i = -n_e e \langle v_i \rangle,
	\end{equation}
	we obtain from (\ref{q_ii}),(\ref{v_ii}) the following relations:

	\begin{equation}
		\label{qrp}
		q_r=-\lambda^{(1)}\frac{dT}{dr},\qquad q_\phi=-B_z\left(\lambda^{(2)}\frac{dT}{dr}\right), \,\,\,q_z =0,
	\end{equation}
	
	\begin{equation}
		\label{jr}
		j_r= e n_e\left(\mu^{(1)}\frac{dT}{dr}\right), \qquad
		j_\phi=en_eB_z\left(\mu^{(2)}\frac{dT}{dr}\right), \,\,\, j_z=0.
	\end{equation}

	The Figs. \ref{cylinder} and \ref{cylinder1} have opposite directions of the initial magnetic field  $B_0$. In both cases this field is deceasing due to the action of the Hall current.
	
	The Lorentz approximation  is applied when the mass of light particles
	(elect\-rons) is much smaller than the the mass of heavy particles (ions or nuclei), and in addition
	electron-electron collisions are neglected. In this approximation the linearized Bol\-tz\-mann equation, from which kinetic coefficients are derived, has an
	exact solution at zero magnetic field.
	In different approaches the solution in Lorentz
	approximation was considered by \cite{Chapman-1952} p. 187,  see also \cite{schatz-1958,BK-2001}.
	
	The explicit exact solution in Lorentz approximation is obtained for the case of a zero magnetic field.
	The heat flux connected only with the temperature gradient, is given in  \cite{schatz-1958}, \cite{BK-2001}
	
	\begin{equation}
		\label{qlor}
		q^T_i=-\tilde\lambda_T \frac{\partial{T}}{\partial x_{\rm i}}, \quad  \tilde\lambda_T
		= \frac{320}{3\pi}\frac{k^2 Tn_e}{m_e}\tau_e.
	\end{equation}
	
	For the average velocity we can write the expression in the Lorentz appro\-xi\-ma\-tion \cite{Glu-2020} with the thermal diffusion for non-degenerate case:
	\begin{equation}
		\label{vlor}
		\langle v_i^{T}\rangle= -\mu^{l}\frac{\partial{T}}{\partial x_{\rm i}}, \quad \mu^{l}\equiv \frac{\sigma_T}{en_e} = \frac{16k}{m_e \pi}\tau_e.
	\end{equation}
	
	Using the expression for the electric current density, we obtain the thermo-diffusion part in the form:
	
	\begin{equation}\label{eq-for-j}
		j_i^T = -n_e e \langle v_i^T \rangle  = \sigma_T \frac{\partial T}{\partial x_i}.
	\end{equation}
	
	We use here parameters:  electron Larmor frequency $\omega_B$, the time between $eN$ collisions $\tau_e$,
	and thermal electrical conductivity coefficient $\sigma_T$, which   in the non-degenerate  Lorentz  gas approximation are determined as
	\cite{BK-2001}
	\begin{eqnarray}
		\label{def}
		\omega_B=\frac{eB}{m_e c}, \quad
		\tau_{e}=
		\dfrac{3}{4}\sqrt{\dfrac{m_e}{2\pi}}\dfrac{(kT)^{3/2}}{Z^2{e}^4n_N\Lambda},\qquad\qquad   \nonumber\\
		\sigma_T = e n_e \mu_{e}=\frac{6\sqrt{2}}{\pi^{3/2}\Lambda}\frac{ e  n_e k^2 T}{n_N  e^4 Z^2}\left(\frac{kT}{m_e}\right)^{1/2}
		= \frac{16k e n_e}{m_e \pi}\tau_e.
		\label{eq3}
	\end{eqnarray}
	Here $n_e,\,\,n_N$ are concentrations of electrons and nuclei with atomic number $Z$,  $\Lambda$ is a Coulomb logarithm.The microscopic process of binary collision is not disturbed here by the magnetic field. For very large magnetic field this approximation is not exact, but it does not change qualitatively the macroscopic behaviour of the system  \cite{Brag-1958a}.
	\\
	\indent Components of the kinetic coefficients tensor in presence of the magnetic field can be expressed using kinetic coefficient  in Lorentz approximation. In particular for thermal electrical conductivity with a $B_z$ magnetic field, the conductivity along magnetic field lines is $\sigma_T$, and across magnetic field lines it is equal to $\sigma_T/(1+\omega_{B}^{2}\tau_{e}^{2})$ In the Hall direction, that is perpendicular to the plane defined by $B_z$ and $\partial T/ \partial x$ the conductivity is written as $\sigma_T \omega_{B} \tau_{e}/(1+\omega_{B}^{2}\tau_{e}^{2})$ \cite{Chapman-1952} p. 322, p. 338.
	Hence  components of the electrical current density vector \textbf{j} in  a cylinder with $B_z$ and temperature gradient vector along the radius is determined as :

	\begin{eqnarray}
		j_r=\frac{\sigma_T ({\bf \nabla}T)_r}{1+\omega_B^2\tau_e^2}, \qquad\qquad\\
		j_\varphi=\frac{(\sigma_T\,({\bf \nabla}T)_r ) \omega_B\tau_e }{1+\omega_B^2\tau_e^2},\quad j_z=0.\nonumber
		\label{eq2}
	\end{eqnarray}

	Connection of vectors ${\textit j_\phi}$ and induced field ${\textit B}$ is determined by Maxwell equations.
	
	\section{Model description, solutions and results}
	%\label{3}
	
	From Maxwell equations we obtain the following relations for the magnetic field components in the cylinder:
	
	\begin{equation}
		B_r=B_\varphi=0,\quad \frac{c}{4\pi}\frac{d B_z}{d r}=-\frac{\sigma_T\,({\bf \nabla}T)_r \omega_B\tau_e }{1+\omega_B^2\tau_e^2}.
		\label{eq5}
	\end{equation}

	\noindent The magnetic field $B_z$ in the cylinder consists of the constant component
	$B_0$, created by external source, and the field $B_{1}$, created by electrical current inside the cylinder.
	
	\begin{equation}
		B_z=B_0+B_1.
		\label{eq6}
	\end{equation}
	
	\noindent Let us consider a stationary state of the cylinder with a constant radial heat flux $Q$. The radial heat flux density is written now as
	
	\begin{equation}
		\label{Q}
		q_r=\frac{Q}{2 \pi r}=-\tilde\lambda_T\frac{(\nabla T)_r}{1+(\omega_B \tau_e)^2}.
	\end{equation}
	This equation should be solved in combination with the equation for $B_z$ written as:
	\begin{equation}
		\frac{d B_z}{d r}=-\frac{4 \pi}{c}\frac{\sigma_T\,({\bf \nabla}T)_r \omega_B\tau_e }{1+(\omega_B\tau_e)^2}.
	\end{equation}
	
	\noindent Using $(\nabla T)_r$  from (\ref{Q}), we obtain the dependencies of the magnetic field  derivative  on the temperature, using (\ref{eq1}), in the form:
	
	\begin{equation}\label{bz}
		\frac{d B_z}{d r} =\frac{3 Q \omega_B \tau_e e}{10  kT c r}.
	\end{equation}
	
	Equations  (\ref{Q}), (\ref{bz})  cannot be extended until the axis with $r = 0$ because of singularities at zero radius. It is suggested in this problem, that the only source of a heat is situated near the axis of the cylinder, and is represented by a uniformly heated cylinder with radius $R_1 << R_0$, $R_0$ is the outer radius of the cylinder.
	
	Equations  (\ref{Q}), (\ref{bz})  are solved jointly under  boundary conditions:
	$B_z(R_0)=B_0$, \quad $T(R_0)=T_0$ at given parameter $Q$. Introducing non-dimensional Hall component $b_1$ as $B_1 = B_0 b_1$, taking into account the definition  $\omega_{B} = \frac{eB_z}{m_e c} = \frac{e(B_0 +B_1)}{m_e c} = \omega_{B0} (1+ b_1)$ and $x = \frac{r}{R_0}$ we  write the Eq. (\ref{bz}) in the form:
	\begin{equation}
		\frac{d b_1}{d x}=  \frac{3 e Q \tau_e }{10 k c T B_{0} x} \omega_{B0} (1+ b_1).
		\label{eq29q}
	\end{equation}
	The Eq.(\ref{Q}) may be written in the following form:
	\begin{equation}
		Q = \frac{-\tilde\lambda_T (\nabla T)_{r} 2 \pi r}{1+\omega^2_{B0} \tau^2_{e} (1+b_1)^{2}}.
		\label{eq30q}
	\end{equation}
	Assuming in (\ref{eq29q}) constant ratio $\tau_e/T = F$,
	the Eq. (\ref{bz})  takes a form:

	\begin{equation}
		\label{eq36}
		\frac{d b_1}{d x}=\frac{3  e Q F}{10 c k    B_0 x }\omega_{B0}(1+b_1), \,\,\, 1> x>x_1  =\frac{R_1}{R_0}, \,\,\, b_1(1) = 0.
	\end{equation}

	Analytical solution of  (\ref{eq36})  is written as:
	
	\begin{equation}
		b_1 =x^{\gamma} - 1, \quad \gamma = \frac{3 e QF}{10 k c B_0}\omega_{B0}.
	\end{equation}

	%The value of $b_1$ is approaching (-1) at $x_1 \to 0$.
	In the case of plasma cylinder with parameters  from (\ref{eq3}), the equations (\ref{eq29q}), (\ref{eq30q}), determining the Hall component $b_1$,  are written as follows:

	\begin{eqnarray}
		\label{syseqforqmodel}
		\frac{d b_1}{d x}= \frac{3  e Q}{10 k c B_0 x } \omega_{B0} (1+ b_1) C_1 T^{1/2}, \nonumber\\
		\frac{dT}{dx}=-\frac{1+ C_1^{2} T^{3} \omega_{B0}^2(1+b_1)^2}{2 \pi x C_2 T^{5/2}} Q .
	\end{eqnarray}
	The constants $C_1$ and $C_2$ are determined from relations:
	\begin{equation}
		\tau _e = 	\frac{3(kT)^{3/2}}{4Z^2 e^4 n_N \Lambda}\sqrt{\frac{m_e}{2 \pi}} = C_1T^{3/2},
	\end{equation}	
	\begin{equation}
		\tilde \lambda_T = \frac{20 kT\sigma_T}{3 e} = \frac{40 \sqrt{2} k n_e}{\pi^{3/2}\Lambda n_N}\left( \frac{kT}{e^2 Z}\right)^{2} \left( \frac{kT}{m_e} \right)^{1/2} = C_2 T^{5/2},
	\end{equation}
	so that:
	
	\begin{equation}
		\omega_B=\omega_{B0}(1+b_1), \quad
		\omega_{B}\tau_e=C_1 T^{3/2}\omega_{B0}(1+b_1).
		\label{eq13a}
	\end{equation}
	Let us introduce dimensionless parameters:
	
	\begin{equation}
		{\tilde T}=\frac{T}{T_0},\,\,	N = \frac{3 e Q\omega_{B0} C_1}{10 kcB_0}T_0^{1/2}, \,\,  G = \ C_1^{2} T_0^3 \omega_{B0}^{2},
		\,\, E=\frac{2\pi C_2 T_0^{7/2}}{Q}.
	\end{equation}	
	Equations (\ref{syseqforqmodel}) have following form with new parameters:
	
	\begin{equation}
		\frac{db_1}{dx} = N\frac{(1+b_1){\tilde T}^{1/2}}{x}, \qquad
		\frac{d\tilde T}{dx}= -\frac{1+G (1+b_1)^2 {\tilde  T^3}}{x E {\tilde T^{5/2}}}.
		\label{syseqforqmodeln}	
	\end{equation}	
	We solve equations (\ref{syseqforqmodeln})  numerically in the interval $x_1\leq x\leq 1$ at boundary conditions:
	
	\begin{equation}
		b_1(1) = 0, \qquad \tilde T(x_1) = 1, \qquad x_1 = 10^{-4}.
	\end{equation}
	
\begin{figure}\label{figureGconst}
		\begin{center}
			\includegraphics[width=0.60\textwidth, angle=270]{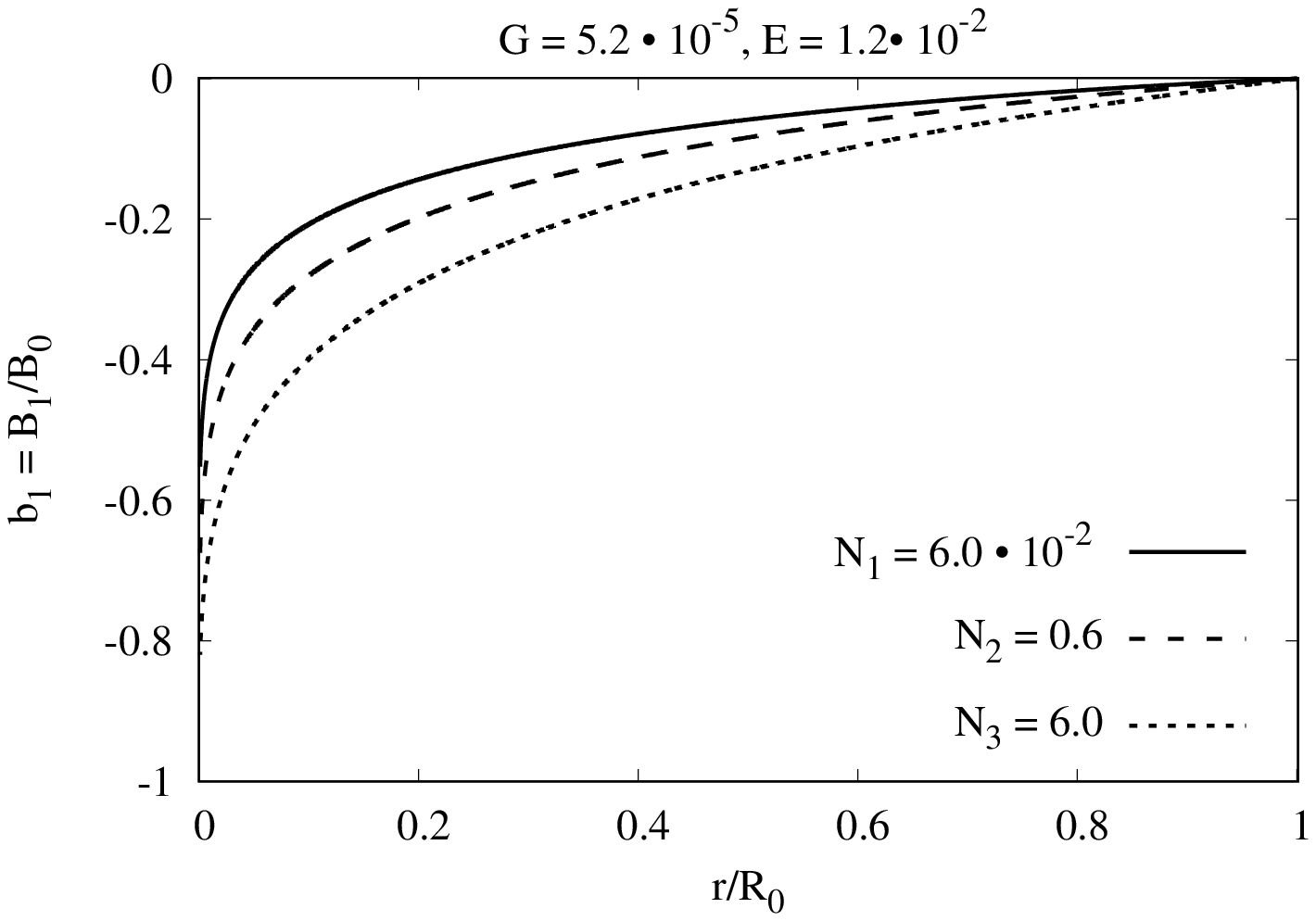}
		\end{center}
		\caption{Magnetic field in the cylinder, induced by the Hall current, for  $G = 5.2\cdot 10^{-5}$, $E = 0.012$, and three values of $N$: $N_1 = 6.0\cdot 10^{-2}$, $N_2 = 0.6$, $N_3 = 6.0$ . These values are related to $Z = 26$, and include combinations
			\\ $B_{0} = 10^{14}\ G , \quad T_0 = 10^{9} \ K,\quad \rho_0 = 10^{9}$ g/cm$^{3}$ for $N_1$
			\\ $B_{0} = 10^{13}\ G , \quad T_0 = 10^{9} \ K,\quad \rho_0 = 10^{8}$ g/cm$^{3}$ for $N_2$;
			\\ $B_{0} = 10^{12}\ G , \quad T_0 = 10^{9} \ K,\quad \rho_0 = 10^{7}$ g/cm$^{3}$ for $N_3$;
			.
		}
	\end{figure}

\begin{figure}
		\begin{center}
			\includegraphics[width=0.60\textwidth, angle=270]{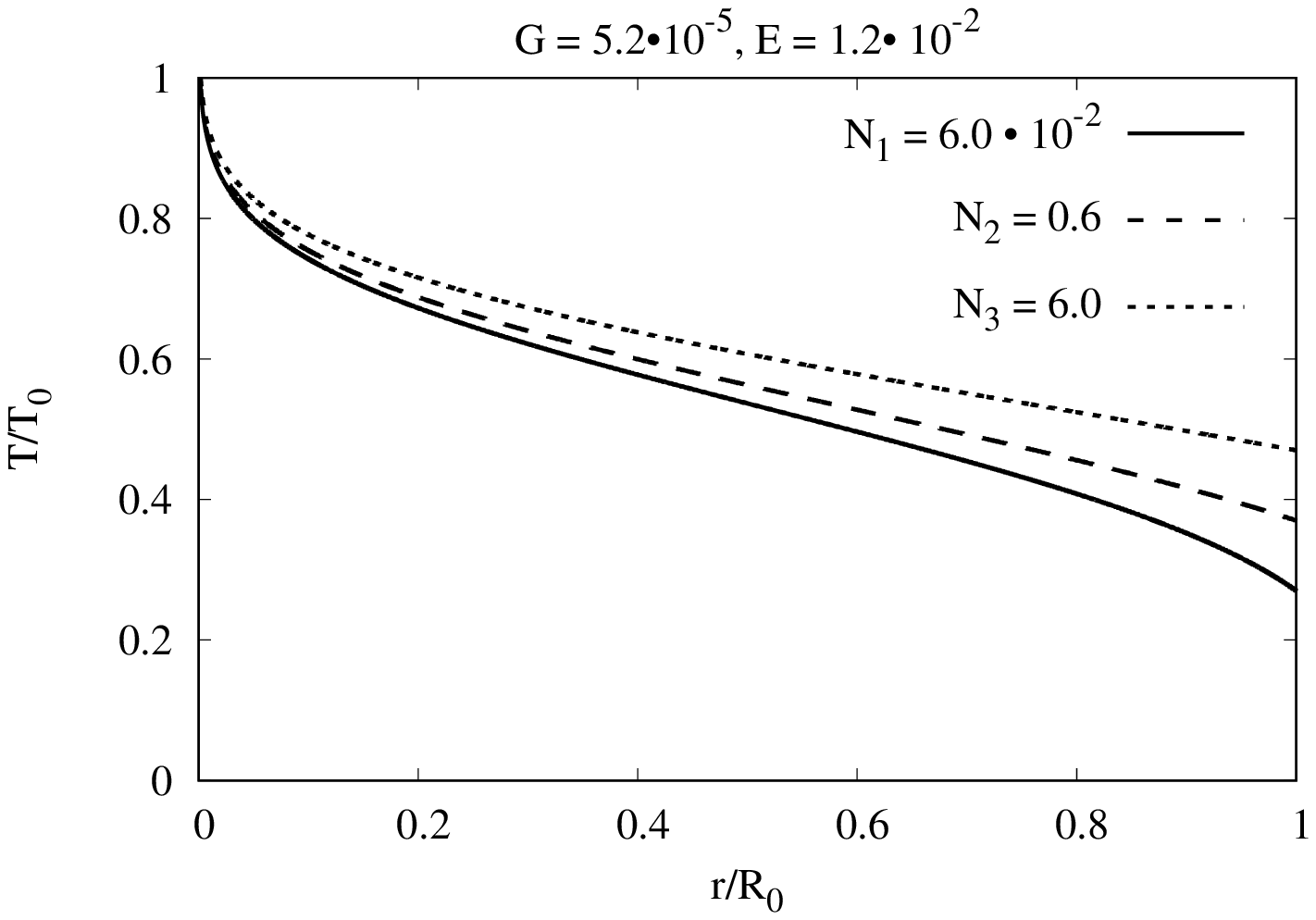}
		\end{center}
		\caption{Temperature distribution in the cylinder for
			the same parameters as in Fig.\ref{figureGconst}.
		}
		\label{figuretempGconst}
	\end{figure}

\begin{figure}
		\begin{center}
			\includegraphics[width=0.60\textwidth, angle=270]{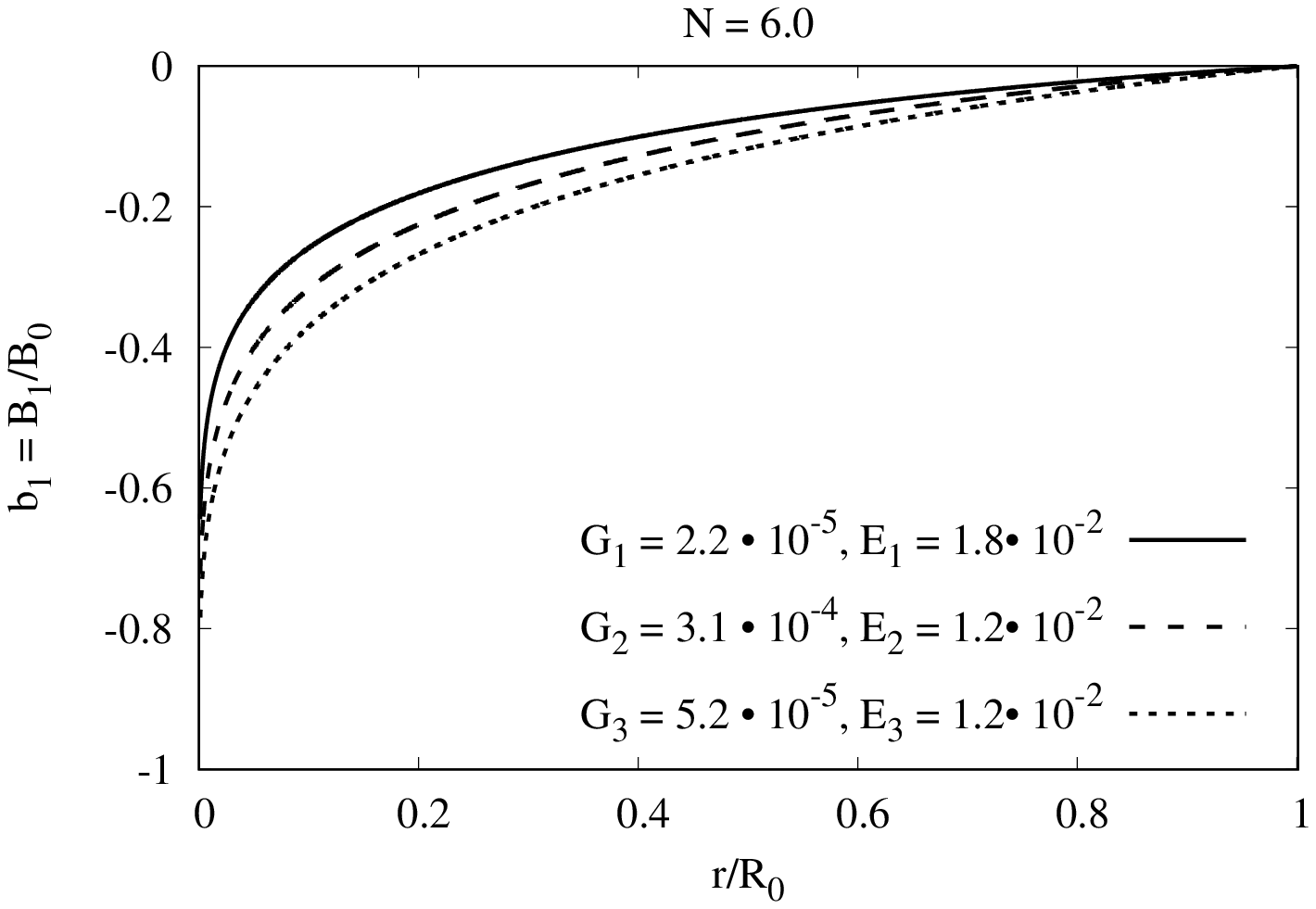}
		\end{center}
		\caption{Magnetic field in the cylinder, induced by the Hall current, for $N = 6.0$  and three variants:
			$G_1 = 2.2\cdot 10^{-5}, E_1 = 0.018$;\,\,\,
			$G_2 = 3.1\cdot 10^{-4}, E_2 = 0.012$;\,\,\,
			$G_3 = 5.2\cdot 10^{-5}, E_3 = 0.012$.
			These values are related to $Z = 26$, and include combinations
			\\$B_{0} = 10^{13}\ G , \quad T_0 = 3.5\cdot 10^{9} \ K, \quad \rho_0 = 10^{9}$  g/cm$^{3}$\quad for $G_1$,$E_1$;
			\\$B_{0} =10^{13} \ G , \quad T_0 = 1.8\cdot 10^{9} \ K, \quad \rho_0 = 10^{8}$ g/cm$^{3}$\quad for $G_2$,$E_2$;
			\\$B_{0} = 10^{12} \ G ,\quad  T_0 = 10^{9} \ K,\quad \rho_0 = 10^{7}$ g/cm$^{3}$\quad  for $G_3$,$E_3$.}
		\label{figureNconst}
	\end{figure}

\begin{figure}
		\begin{center}
			\includegraphics[width=0.60\textwidth, angle=270]{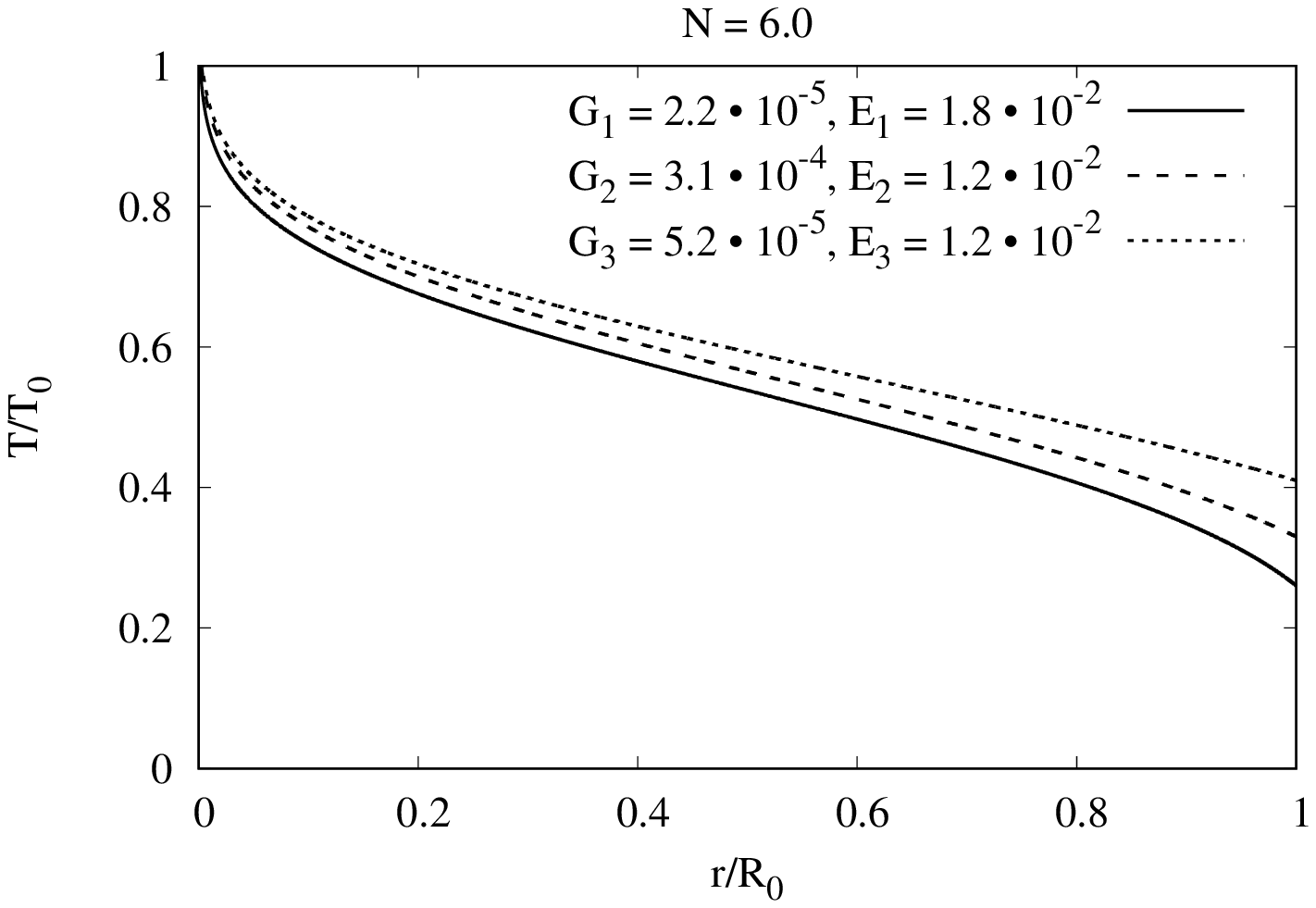}
		\end{center}
		\caption{Temperature distribution in the cylinder for
			the same parameters as in Fig.\ref{figureNconst}}
		\label{figuretempNconst}
	\end{figure}

\begin{figure}
		\begin{center}
			\includegraphics[width=0.60\textwidth, angle=270]{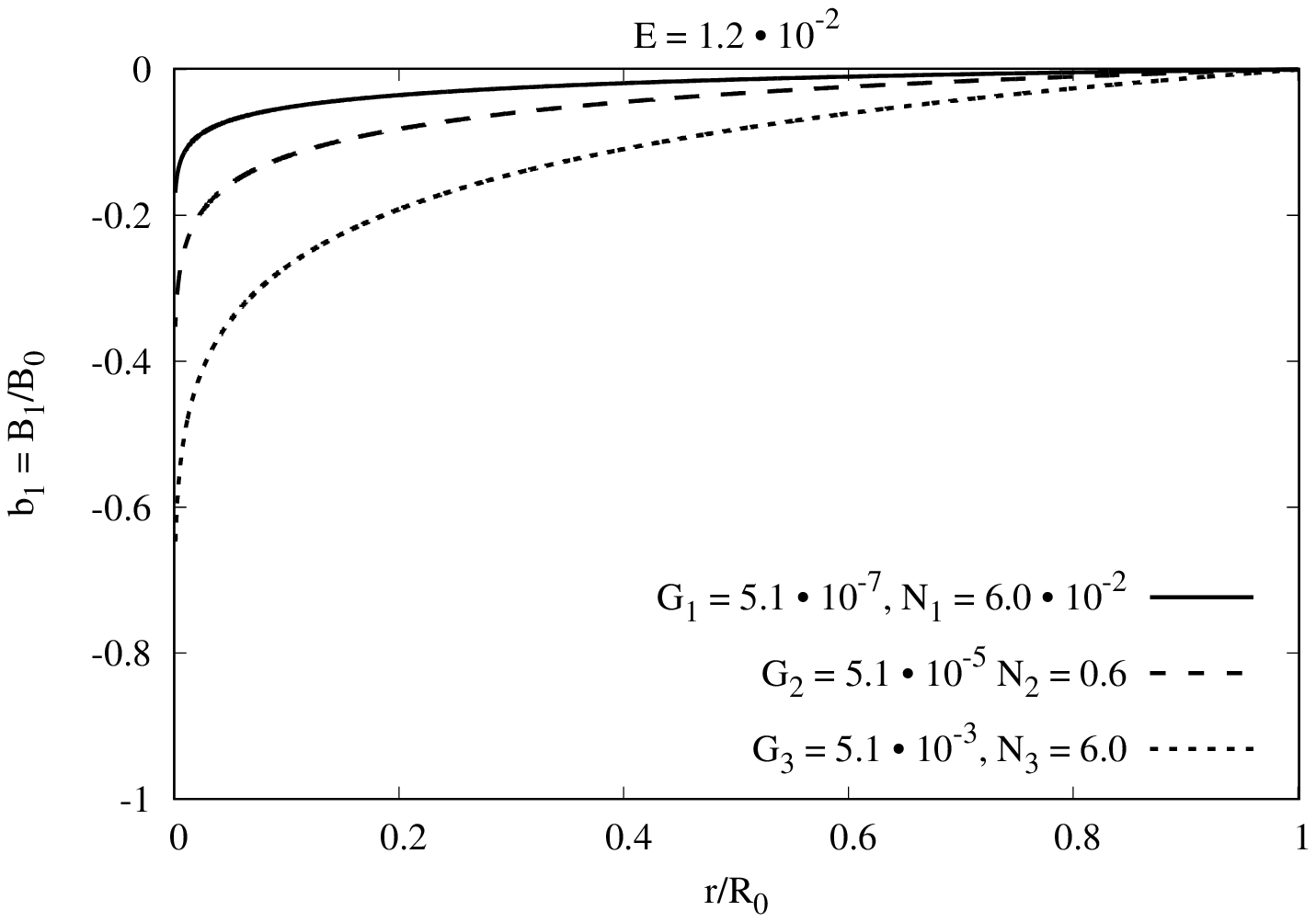}
		\end{center}
		\caption{Magnetic field in the cylinder, induced by the Hall current, for $E=0.012$, and  three variants:
			$G_1=5.1\cdot 10^{-7}, N_1 =0.06$.
			$G_2=5.1\cdot 10^{-5},N_2=0.6$;\,\,\,
			$G_3=5.1\cdot 10^{-3},N_3=6.0$;\,\,\,
			These values are related to $Z = 26$, and include combinations
			\\$B_{0} = 10^{13}\, G , \quad T_0 =10^{9}\ K, \quad \rho_0 = 10^{9}$\, g/cm$^{3}$ \quad for $G_1$,$N_1$;
			\\$B_{0} =10^{13}\, G, \quad T_0 =10^{9}\, K, \quad \rho_0 = 10^{8}$\, g/cm$^{3}$ \quad for $G_2$,$N_2$;
			\\$B_{0} = 10^{13}\, G , \quad T_0 = 10^{9}\, K, \quad \rho_0 = 10^{7}$\, g/cm$^{3}$ \quad  for $G_3$, $N_3$.
			  }
		\label{figureEconst}
	\end{figure}

\begin{figure}
		\begin{center}
			\includegraphics[width=0.60\textwidth, angle=270]{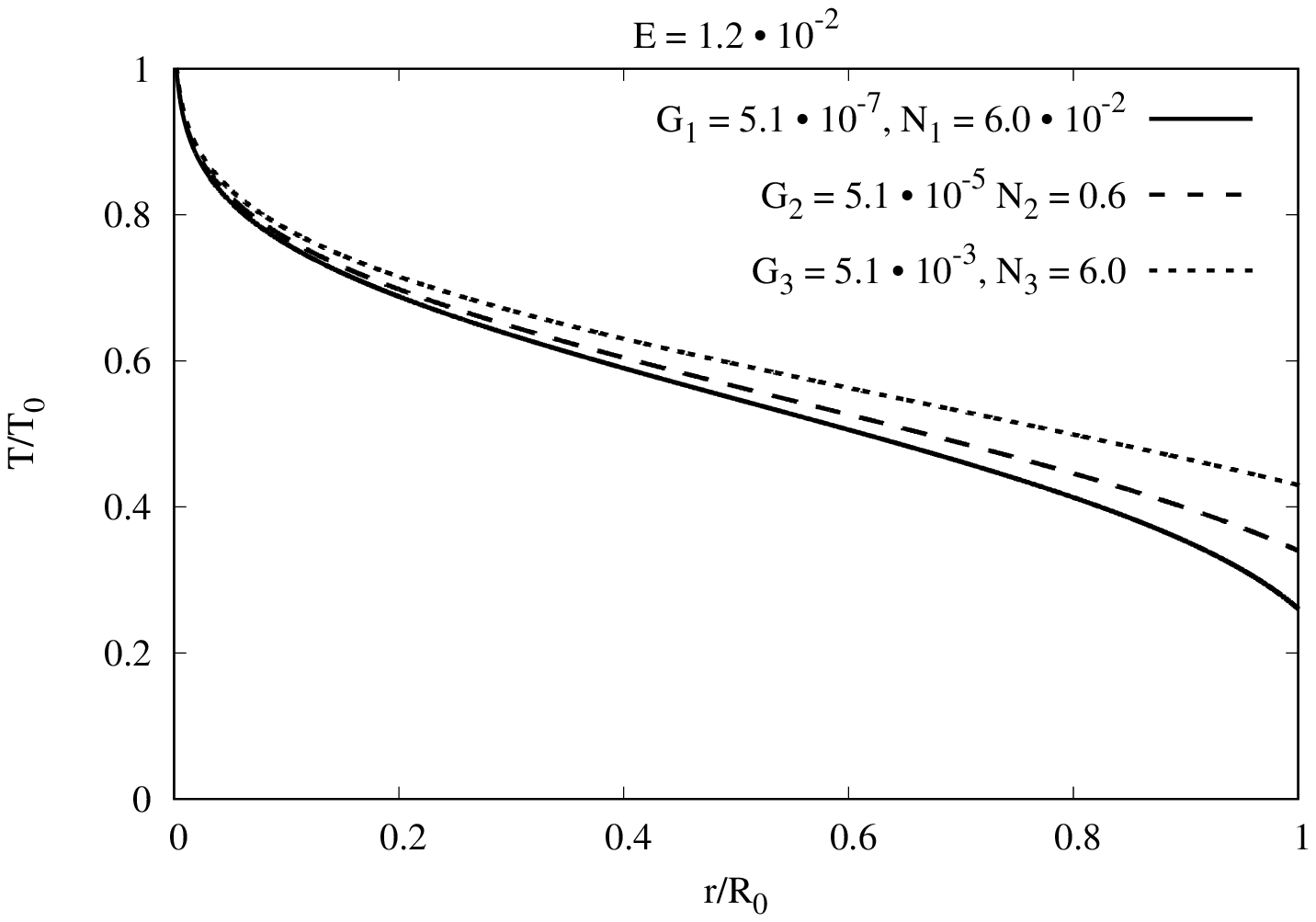}
		\end{center}
		\caption{Temperature distribution in the cylinder for
			the same parameters as in Fig.\ref{figureEconst}
		}\label{figuretempEconst}
	\end{figure}

	Results of the solution are presented on the figures (\ref{figureGconst}) - (\ref{figuretempEconst}) for the case of plasma parameters in the neutron star crust.
	
	\begin{figure}
		\begin{center}	\includegraphics[width=0.60\textwidth, angle=270]{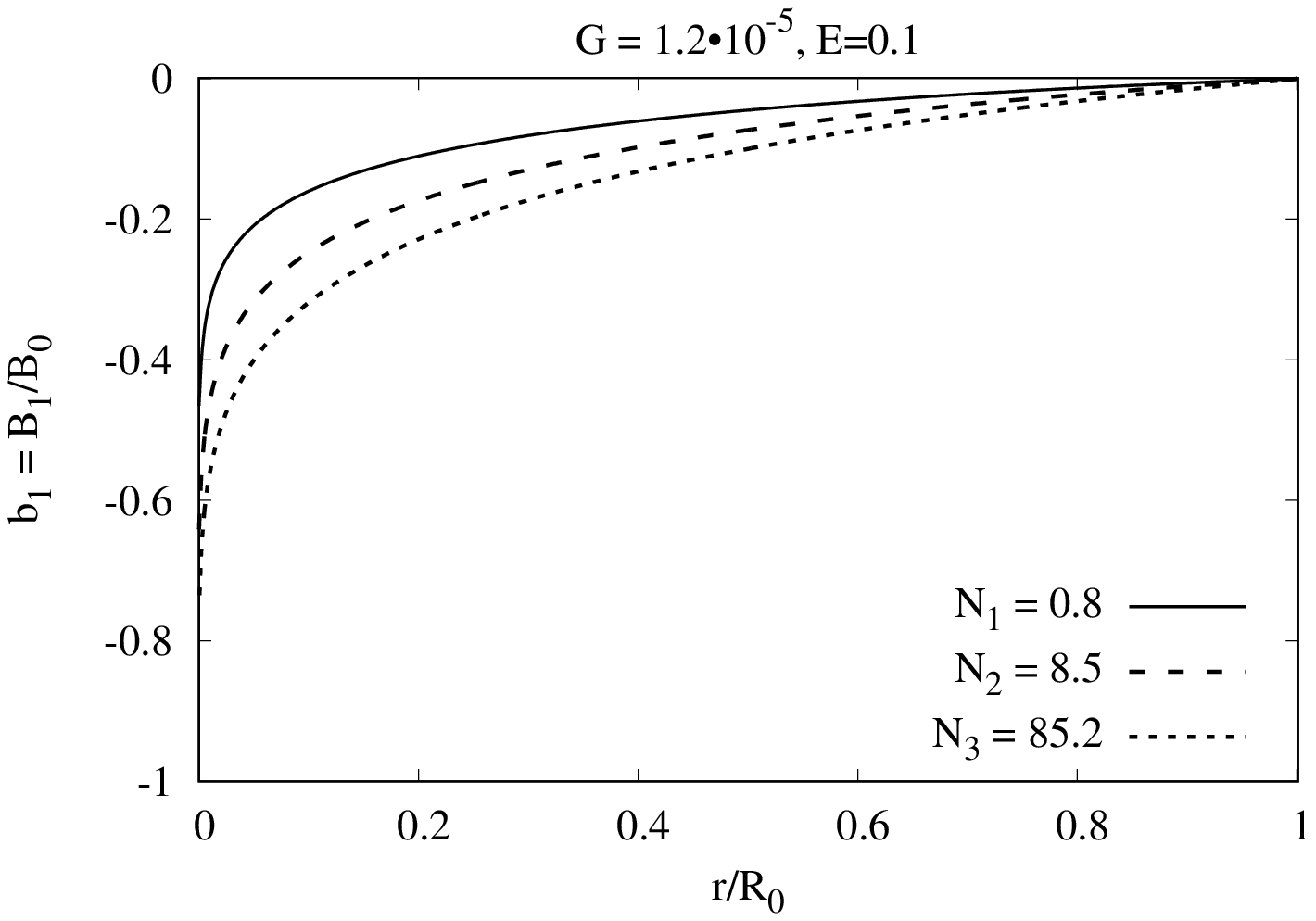}
		\end{center}
		\caption{Magnetic field in the cylinder, induced by the Hall current, for $G=1.2\cdot 10^{-5}$, $E=0.1$,  and three variants: $N=0.8$; $N_2=8.5$;  $N_3=85.2$. These values are related to $Z =1$ and include combinations
			\\$B_{0} = 5\cdot 10^3$ G, $T_0 =2\cdot 10^{5}$ K,
			$\rho_0=10^{-4}$ g/cm$^{3}$  for $N_1$;
			\\$B_{0}=5\cdot 10^{2}$ G,  $T_0 = 2\cdot 10^{5}$ K, $\rho_0 = 10^{-5}$ g/cm$^{3}$  for $N_2$;
			\\$B_{0}=50$ G, $T_0 = 2\cdot 10^{5}$ K, $\rho_0=10^{-6}$ g/cm$^{3}$ for $N_3$. }
		\label{figureGconstlab}
	\end{figure}
	
	\begin{figure}
		\begin{center}
			\includegraphics[width=0.60\textwidth, angle=270]{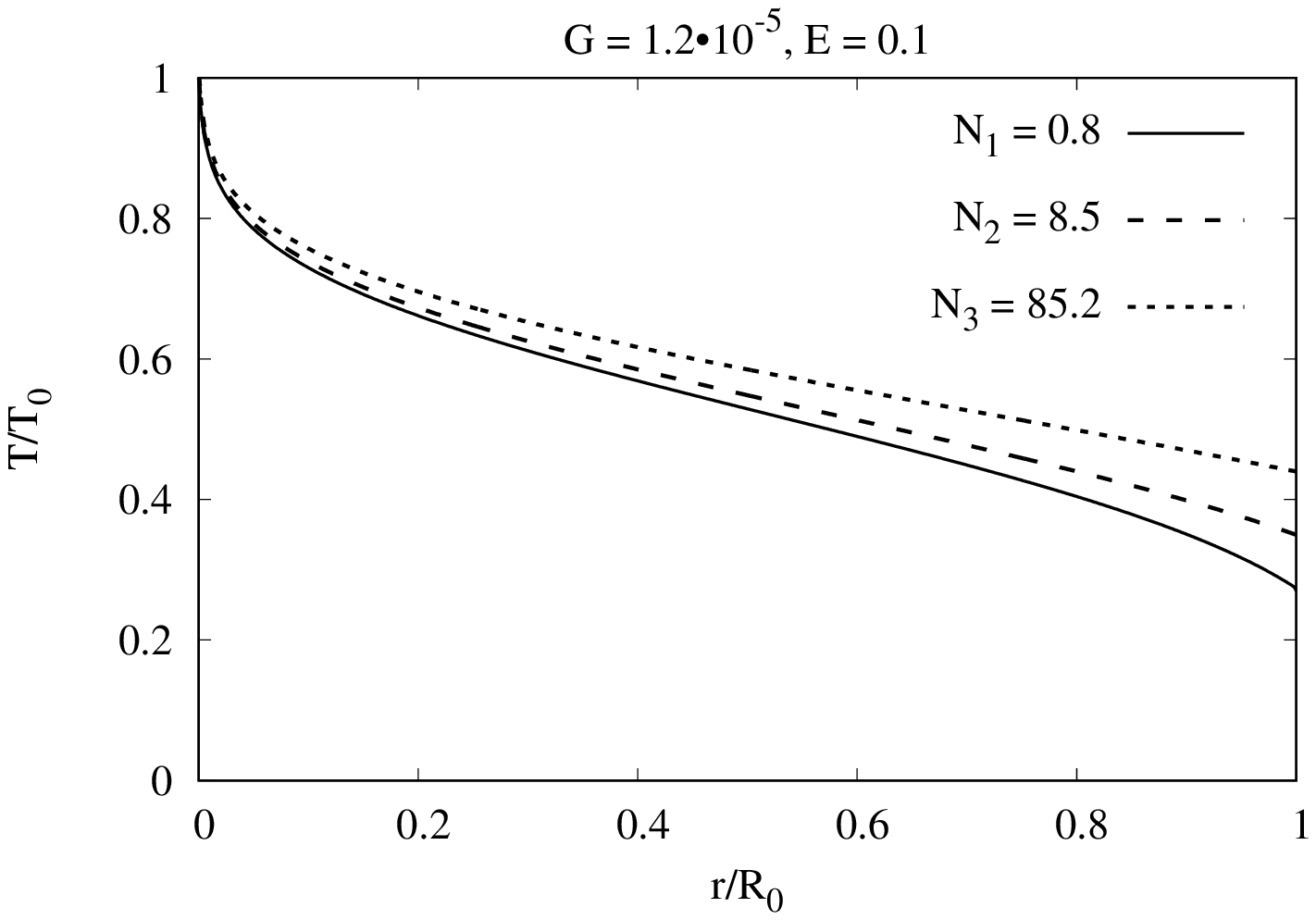}
		\end{center}
		\caption{Temperature distribution in the cylinder for
			the same parameters as in Fig.\ref{figureGconstlab}.
		}
		\label{figuretempGconstlab}
	\end{figure}
	
	\begin{figure}
		\begin{center}
			\includegraphics[width=0.60\textwidth, angle=270]{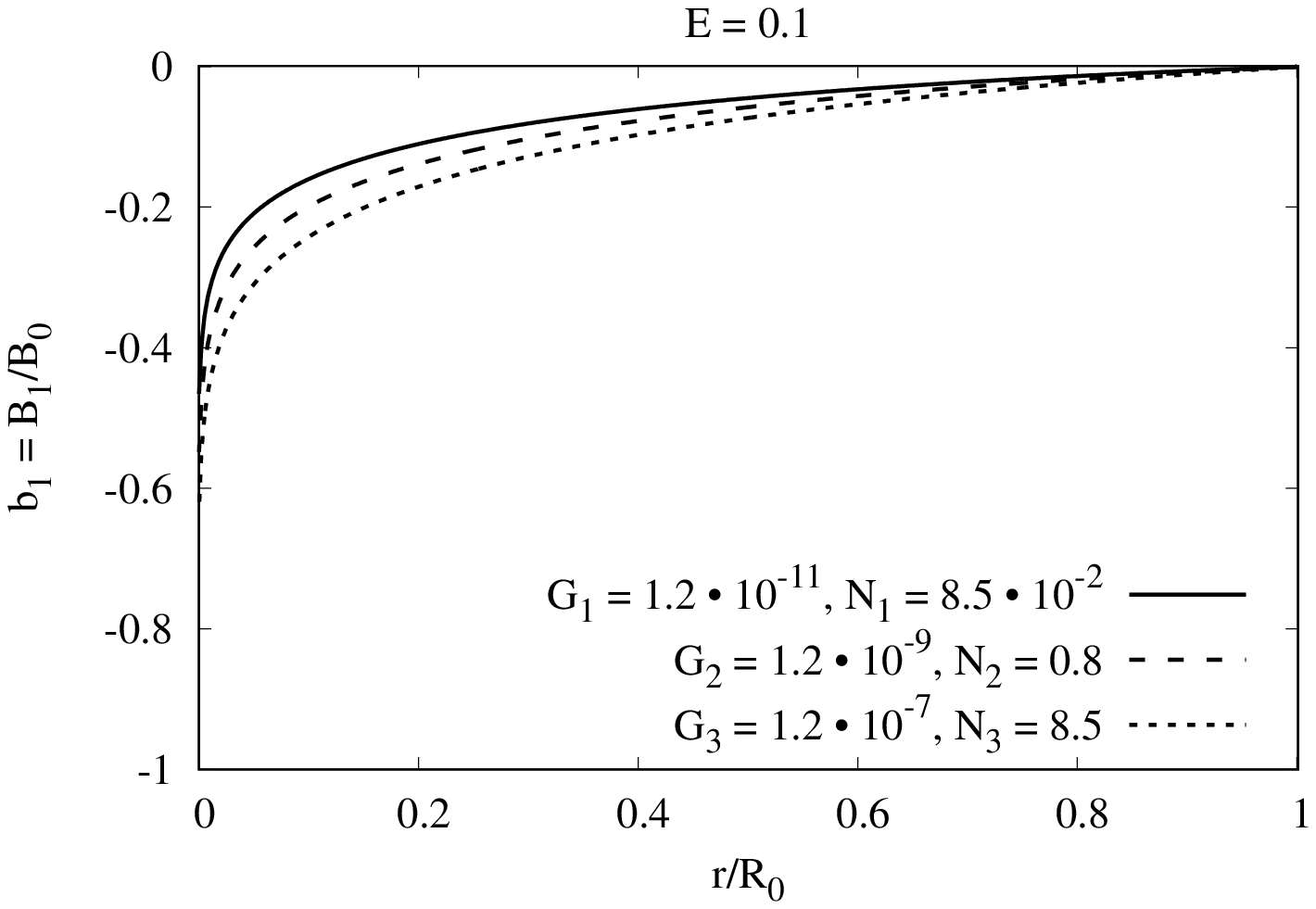}
		\end{center}
		\caption{Magnetic field in the cylinder, induced by the Hall current, for $E = 0.1$ and three variants: $G_1= 1.3\cdot 10^{-11}$, $N_1=0.085$;  $G_2=1.3\cdot 10^{-9}$,  $N_2=0.8$;
			$G_3=1.3\cdot 10^{-7}$, $N_3=8.5$. These values are related to $Z = 1$, and include variants
			$T_0 = 2\cdot 10^{5}$ K, $B_{0}= 50$ G ,\\
			$\rho_0=10^{-3}$ g/cm$^{3}$ for $N_1$,$G_1$; \qquad
			$\rho_0 = 10^{-4}$ g/cm$^{3}$  for $N_2$,$G_2$; \qquad
			$\rho_0 = 10^{-5}$ g/cm$^{3}$  for $N_3$ $G_3$. 	 }
		\label{figureEconstlab}
	\end{figure}
	
	\begin{figure}
		\begin{center}
			\includegraphics[width=0.60\textwidth, angle=270]{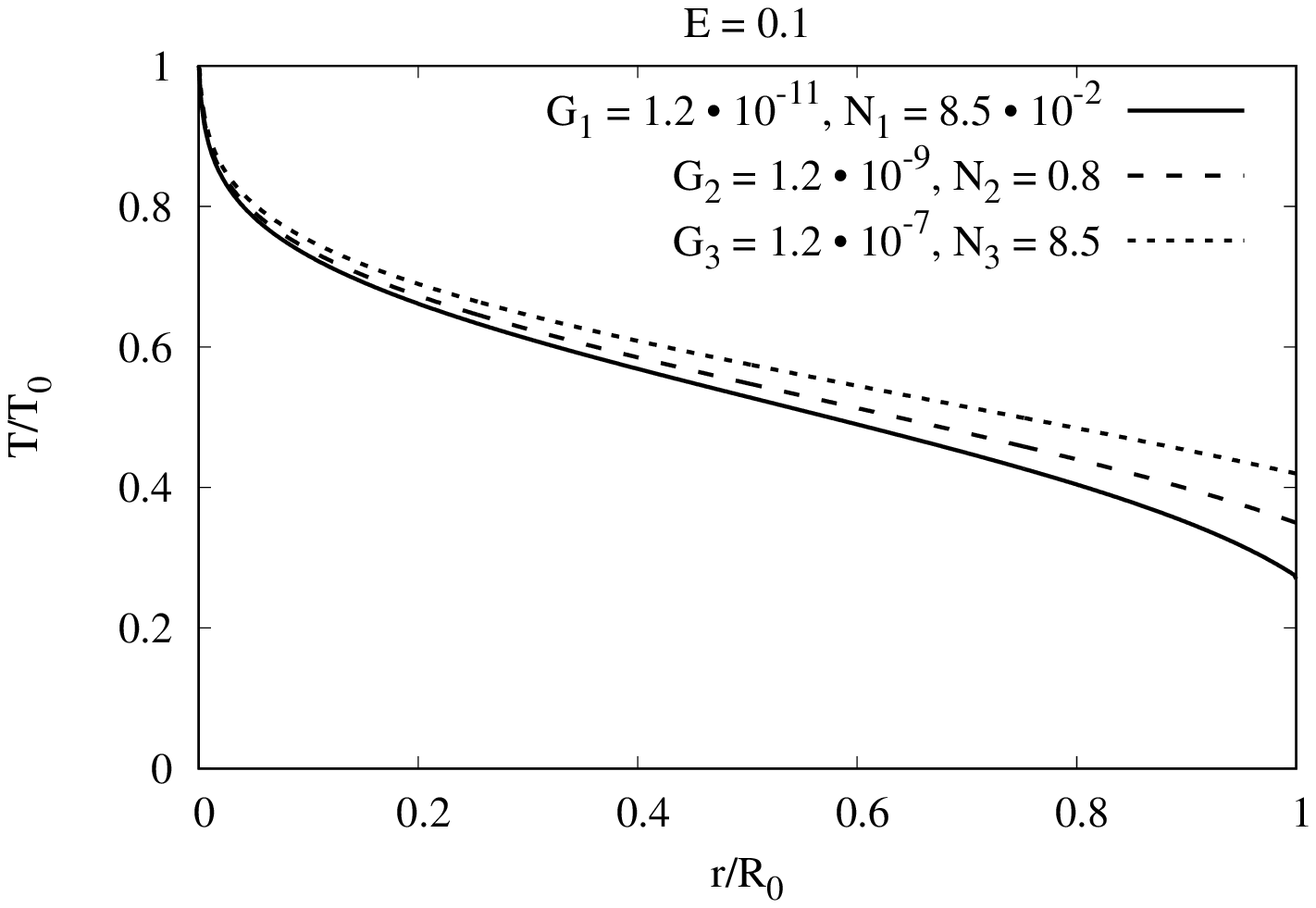}
		\end{center}
		\caption{Temperature distribution in the cylinder for
			the same parameters as in Fig.\ref{figureEconstlab}.
		}\label{figuretempEconstlab}
	\end{figure}
	\pagebreak
	\begin{figure}
		\begin{center}
			\includegraphics[width=0.60\textwidth, angle=270]{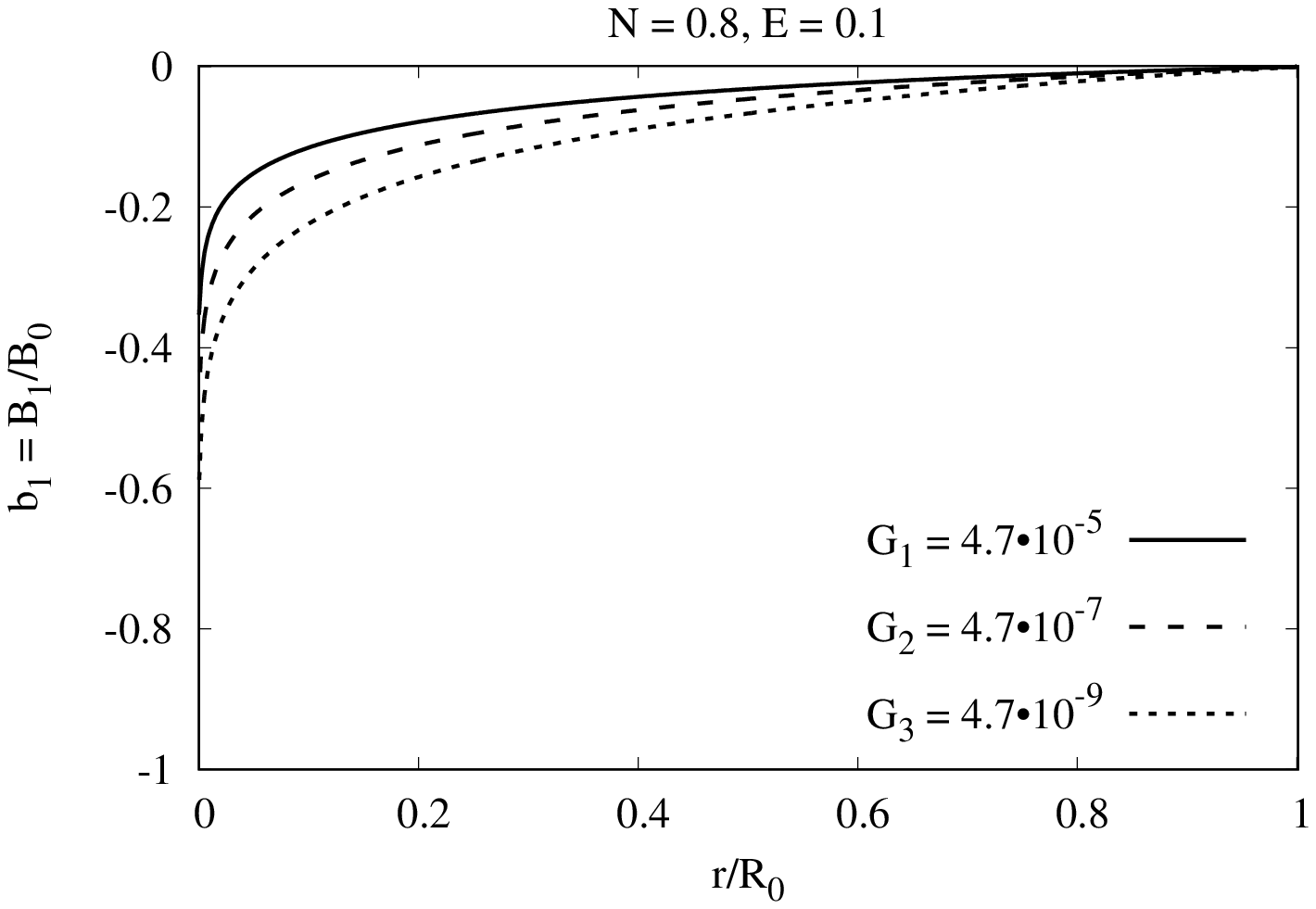}
		\end{center}
		\caption{Magnetic field in the cylinder, induced by the Hall current, $N = 0.8,\,\,E=0.1$,  and three variants: $G_1 = 4.7\cdot 10^{-5}$; $G_2 = 4.7\cdot 10^{-7}$; $G_3 = 4.7\cdot 10^{-9}$. These values are related to $Z = 1$, and include variants $\rho=10^{-4}$ g/cm$^{3}$, $T_0 = 2\cdot 10^{5}  \, K $,
			\\$B_{0} = 10^{4}$G, for $G_1$;\qquad
			$B_{0} =10^{3}$ G, for $G_2$; \qquad
			$B_{0} =10^{2}$ G, for $G_3$.}
		\label{figureNconstlab}
	\end{figure}
	
	\begin{figure}
		\begin{center}
			\includegraphics[width=0.60\textwidth, angle=270]{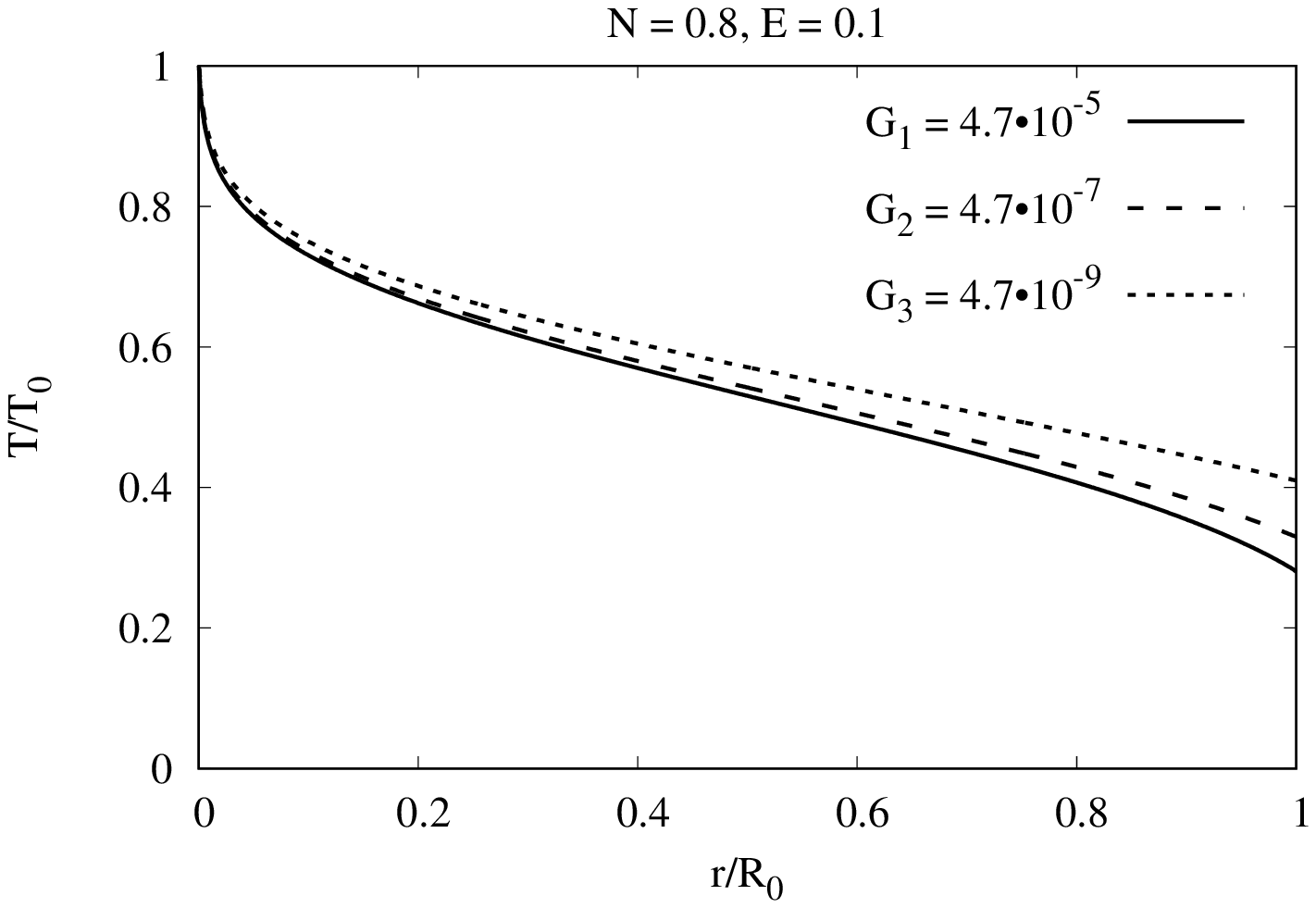}
		\end{center}
		\caption{Temperature distribution in the cylinder for
			the same parameters as in Fig.\ref{figureNconstlab}.
		}
		\label{figuretempNconstlab}
	\end{figure}
	
	\noindent Equations (\ref{syseqforqmodeln}) can be used for analyzing the  magnetized plasma in laboratory facilities. Results of these calculations are presented on the figures (\ref{figureGconstlab}) - (\ref{figuretempNconstlab}).
	\bigskip
	\bigskip
	\pagebreak
	
	\section{Discussion}
	
	It is shown in this paper that the magnetic field, generated by
	the azimuthal Hall current, decreases the magnetic field,
	produced by external sources. Equation, determining $B_1/B_0$ ratio
	of the magnetic filed produced by the Hall current to the external
	magnetic field, is derived.
	Hall current in the present consideration is produced by
	temperature gradient for the case when diffusion vector is equal to
	zero.
	Analytical results are obtained for the case, when coefficients of
	heat conductivity, electroconductivity, and a time between
	collisions are constant. Results of numerical calculations performed for the
	case of plasma parameters in neutron star envelopes, are shown in figures
	(\ref{figureGconst})-(\ref{figuretempNconstlab}).
	The calculations  for parameters, related to
	laboratory plasma, are presented in figures
	(\ref{figureGconstlab})-(\ref{figuretempNconstlab}).
	
	Kinetic coefficients in the magnetic field are determined by
	tensors, connected with temperature gradient and diffusion vector.
	Influence of the Hall current on the temperature
	distribution, structure of magnetic and electric fields,
	in realistic geometry of neutron star envelope needs
	further consideration. It can be  important for  modeling of the structure of the magnetic field along the surface of the neutron star, and for studying a coupled magneto-thermal evolution  of temperature, magnetic and electric fields in neutron stars.	
	The electrons in the inner envelope of the neutron star
	may become degenerate and relativistic in conditions of
	high density and temperature. We have used non-relativistic and non-degenerate approximation for transport coefficients in all our calculations. Therefore the results presented in Figs.3-8 can be considered as correct only qualitatively. Account of relativistic corrections and degeneracy in calculations of transport coefficients of plasma meets with difficulties, so analytical formulae for these conditions have been obtained approximately, with considerable simplifications. In the situation, when the structure of the neutron star is far from a very simple   cylindrical model, used here, we have done calculations of the non-linear Hall effects using simplified transport coefficients for neutron star parameters. \\
	\indent In recent years  experimental  study of astrophysical processes is developing (laboratory astrophysics).  The goal is  to model astrophysical processes in terrestrial laboratory, basing on the similarity theory relations. Our results can be useful for   studying the Hall current effects in the  laboratory plasma, which may be applied for astrophysical conditions. High temperature gradients in presence of very strong  magnetic fields are formed during stellar core collapses, leading to formation of neutron stars, accompanying by supernovae explosions. The new born neutron star is very hot, strongly magnetized, and with large temperature gradients. Thermoelectric processes are very important on this short (few years) stage of the neutron star life, during a rapid cooling by neutrino energy losses \cite{Tsu-1965}.
	The magnetic field structure formed in this short stage kips frozen, and the time of its slow changes may exceed  millions of years.

	\makeatletter
	\def\fps@table{h}
	\def\fps@figure{h}
	\makeatother

\end{document}